# Hypergeometric Functions of Nilpotent Operators: Functional Collapse and Structural Depth at Exceptional Points

**Author:** Ramón Moya

**Affiliation:** School of Mathematics, Universidad Autónoma de Santo Domingo (UASD), Dominican Republic

**Email:** rmoya07@uasd.edu.do ·

**ORCID:** 0009-0001-1601-4699

**Abstract**

We study hypergeometric functions of nilpotent operators in finite-dimensional settings, motivated by the algebraic structure of exceptional points in non-Hermitian quantum mechanics. Our starting point is the following exact result: if $N$ is a nilpotent operator of index $m + 1$ in an associative algebra over $\mathbb{C}$, then every generalized hypergeometric function $_pF_q$ evaluated at $N$ reduces to a finite polynomial in $N$ of degree at most $m$, without any analytic convergence requirement. This "functional collapse" is distinct from the classical parameter-termination mechanism and arises purely from the nilpotent structure of the argument.

The main result is a "nilpotent depth criterion" (Theorem 2): if the first non-constant coefficient of a formal series $F$ appears in degree $r \geq 1$, then the nilpotent part $F(N) - F(0)I$ has nilpotency index bounded above by $\lceil (m + 1)/r \rceil$. This bound is sharp in generic cases and provides a quantitative measure of how many Jordan levels survive after applying a special function.

We apply this criterion to Hamiltonians at exceptional points, where $H = \lambda I + N$ with $N^{m+1} = 0$. Theorem 3 establishes that a function $F$ analytic at $\lambda$ reduces the Jordan depth of the exceptional point from $m + 1$ to at most $\lceil (m + 1)/r \rceil$, where r is the contact order of F at λ (the order of the first nonzero term in the Taylor expansion of $F(z) - F(\lambda)$ around $z = \lambda$). As consequences: the time evolution operator $e^{tH}$ preserves the full Jordan depth for all $t \neq 0$; a function with a zero of order $m + 1$ at $\lambda$ annihilates the entire Jordan structure; and the order of the pole of the modified resolvent is reduced from order $m + 1$ to at most order $m + 1 - r$. Results are illustrated with explicit $3 \times 3$ Jordan block computations for $_1F_1$, $_2F_1$, and the time evolution operator, confirming sharpness of the bounds.

**Quick reference table**

| Input | Output | Mechanism |
|---|---|---|
| $N^{m+1} = 0, F(z) = \sum c_j z^j$ | $F(N) = \sum_{j=0}^{m} c_j N^j$ (finite sum) | Lemma 1 |
| $F$ of order $r$ at $0$, $N^{m+1} = 0$ | nilpotency index of $F(N) - F(0)I \leq \lceil (m+1)/r \rceil$ | Theorem 2 |
| $H = \lambda I + N$, $F$ of contact order $r$ at $\lambda$ | Jordan depth of $F(H)$ reduced to $\leq \lceil (m+1)/r \rceil$ | Theorem 3 |
| Any $N \in M_n(\mathbb{C})$ nilpotent | $\operatorname{tr}(_pF_q(a; b; N)) = n$ | Corollary 2 |

## 1. Introduction

The generalized hypergeometric function is given by

$$_pF_q\left(\begin{matrix}a_1, \ldots, a_p \\ b_1, \ldots, b_q\end{matrix}; z\right) = \sum_{j=0}^{\infty} \frac{(a_1)_j \cdots (a_p)_j}{(b_1)_j \cdots (b_q)_j} \frac{z^j}{j!},$$

where $(a)_j$ denotes the Pochhammer symbol: $(a)_0 = 1$ and $(a)_j = a(a+1)\cdots(a+j-1)$ for $j \geq 1$.

Classically, the series terminates when one of the upper parameters is a non-positive integer. If $a_1 = -m$ with $m \in \mathbb{N}$, then $(-m)_j = 0$ for all $j \geq m + 1$, and the series reduces to a polynomial of degree $m$ in $z$. This is the classical "parameter-termination" mechanism.

The goal of this work is to separate this from a distinct and independent mechanism: "nilpotent-argument termination". If the argument is not an ordinary complex number but a nilpotent element $N$ of a $\mathbb{C}$-associative algebra with unit, the series becomes finite because the powers $N^j$ vanish beyond a certain index, even if no parameter is a negative integer.

Beyond the collapse itself, we study a finer structural property: how many Jordan levels survive after applying $F$? If the first $r - 1$ non-constant coefficients of $F$ vanish, the nilpotent part $F(N) - F(0)I$ begins at $N^r$, and its nilpotency index is bounded by $\lceil (m + 1)/r \rceil$. This "nilpotent depth criterion" (Theorem 2) is the main algebraic result.

We then apply these results to Hamiltonians with exceptional points (Theorem 3), where the nilpotent structure is not an abstraction but the literal algebraic content of the physics. Exceptional points of order $m + 1$ are, by definition, operators of the form $H = \lambda I + N$ with $N^{m+1} = 0$; the entire Jordan structure of the exceptional point is controlled by the nilpotent part $N$.

## 1.1 Relation to prior literature

The evaluation of analytic functions on nilpotent matrices is a classical topic of matrix functional calculus and Jordan theory; see Horn–Johnson [3] and Higham [7] for comprehensive treatments. In that setting, a holomorphic function $f$ evaluated at a nilpotent matrix $N$ yields a finite polynomial expression derived from the Taylor expansion of $f$ around zero, since all eigenvalues of $N$ are zero.

Classical hypergeometric theory studies termination due to negative integer upper parameters, integer parameter differences, and special reduction formulas, including Karlsson–Minton type formulas; see Andrews–Askey–Roy [1] and Bailey [2].

The theory of exceptional points in non-Hermitian quantum mechanics is a growing field; foundational references include Kato [8], Heiss [9], and Bender–Boettcher [10]. The algebraic structure of exceptional points — as Jordan blocks — is well understood, but the behavior of special functions evaluated at such operators does not appear to have been studied systematically.

The present work connects these three areas. It does not introduce new classical hypergeometric identities. Its goal is to develop a framework for interpreting algebraic truncations induced by nilpotency, to quantify the Jordan depth of $F(N)$ via the nilpotent depth criterion, and to derive consequences for the spectral theory of Hamiltonians at exceptional points. To the best of our knowledge, this perspective has not been developed explicitly in this form in the standard literature.

## 1.2 Motivation from mathematical physics

Special functions of operators appear throughout mathematical physics. Classical examples include exponentials of Hamiltonians, resolvents, propagators, and hypergeometric functions associated to special differential equations.

When the operator is diagonalizable, functional calculus reduces to evaluating the function on the spectrum. When the operator contains nilpotent components, however, the situation changes qualitatively: Jordan blocks appear and produce additional polynomial terms in the spectral expansion.

This paper focuses precisely on that case. The nilpotency induces an exact algebraic truncation, allowing hypergeometric functions of operators to be evaluated through finite sums. This phenomenon is natural in contexts involving defective operators, non-diagonalizable Hamiltonians, exceptional points in PT-symmetric systems, and finite-dimensional quantum evolutions.

**Concrete example (PT-symmetric 2×2 system).** Consider the gain-loss Hamiltonian

$$H = \begin{pmatrix} \omega - i\gamma & \kappa \\ \kappa & \omega + i\gamma \end{pmatrix}.$$

At the exceptional point $\kappa = \gamma$, the eigenvalues coalesce to $\lambda = \omega$ and $H = \omega I + N$ with

$$N = \begin{pmatrix} -i\gamma & \gamma \\ \gamma & i\gamma \end{pmatrix}, N^2 = 0.$$

The time evolution is $e^{tH} = e^{\omega t}(I + tN)$. The linear growth $\gamma t\, e^{\omega t}$ in the off-diagonal entries is the observable signature of the exceptional point of order 2. The dynamical passage through such an exceptional point and its physical consequences have been analyzed in [13]. Theorem 3 will show that any function F with F'(ω)≠0 preserves this signature, while a function with contact order $r = 2$ at $\omega$ would annihilate the nilpotent structure entirely.

The organization is as follows. Section 2 establishes the algebraic framework. Section 3 proves the fundamental termination lemma. Section 4 applies it to $_pF_q$. Section 5 separates the two finiteness mechanisms. Section 6 studies structural properties of $F(N)$. Section 7 analyzes the matrix case. Section 8 states and proves the nilpotent depth criterion with a corollary on composition. Section 9 interprets the criterion. Section 10 develops four explicit examples. Section 11 applies the framework to exceptional points in non-Hermitian Hamiltonians, with explicit $3 \times 3$ computations. Section 12 discusses Karlsson–Minton formulas as future direction. Section 13 summarizes the contribution and limitations.

## 2. Algebraic framework

Let $\mathcal{A}$ be a $\mathbb{C}$-associative algebra with unit $I$. An element $N \in \mathcal{A}$ is "nilpotent" if there exists $s \in \mathbb{N}$ with $N^s = 0$. If $N^{m+1} = 0$ and $N^m \neq 0$, we say $N$ has "nilpotency index exactly" $m + 1$.

When $N^{m+1} = 0$, the commutative subring generated by $N$ is isomorphic to the truncated algebra:

$\mathbb{C}[N] \cong \mathbb{C}[x]/(x^{m+1})$.

In that quotient, every polynomial reduces to degree at most $m$, and every formal series in $N$ collapses to a finite sum. This is the algebraic foundation of all results in the paper.

**Observation 2.1 (Algebraic perspective).** From an algebraic perspective, the assignment $F(z) \mapsto F(N)$ defines a $\mathbb{C}$-algebra homomorphism

$\text{ev}_N: \mathbb{C}[[z]] \longrightarrow \mathbb{C}[N] \cong \mathbb{C}[x]/(x^{m+1})$,

where $\mathbb{C}[[z]]$ denotes the ring of formal power series. Theorem 2 describes how this evaluation homomorphism interacts with the natural filtration of $\mathbb{C}[[z]]$ by order of vanishing at the origin.

**Observation 2.2 (Connection with the SON framework).** The algebra $\mathbb{C}[N] \cong \mathbb{C}[x]/(x^{m+1})$ is the same truncated power series ring $V_n = \mathbb{C}[[\varepsilon]]/(\varepsilon^{n+1})$ underlying the Unified Nilpotent Operational System (SON) of [6]. In that framework, nilpotency of index $m + 1$ converts superexponential classical algorithms into polynomial-length computations via Newton iteration. The present paper can be seen as a functional-calculus counterpart of this principle, in the specific context of hypergeometric functions of nilpotent operators: nilpotency of the argument converts every infinite formal series into a polynomial of degree at most $m$, without any convergence requirement. The nilpotent depth criterion (Theorem 2) is the analogue, in the setting of special functions, of the complexity bounds established in [6].

**Canonical examples.**

(a) The Jordan matrix $N = J_n(0) \in M_n(\mathbb{C})$, with ones on the superdiagonal, satisfies $N^n = 0$ and $N^{n-1} \neq 0$: nilpotency index $n$. This is the standard model of an exceptional point of order $n$.

(b) The truncated differentiation operator $D = d/dx$ on $\mathcal{P}_m$ satisfies $D^{m+1} = 0$: index $m + 1$.

(c) The element $\varepsilon = x + (x^{m+1})$ in $\mathbb{C}[x]/(x^{m+1})$ is the universal model of a nilpotent of index $m + 1$.

## 3. Fundamental termination lemma

**Lemma 1 (Exact termination by nilpotency).** Let $\mathcal{A}$ be a $\mathbb{C}$-associative algebra with unit. Let $N \in \mathcal{A}$ with $N^{m+1} = 0$. Let $F(z) = \sum_{j=0}^{\infty} c_j z^j$ be a formal power series with $c_j \in \mathbb{C}$. Then

$$F(N) = \sum_{j=0}^{m} c_j N^j.$$

The identity is algebraic and requires no convergence hypothesis on $F(z)$.

**Proof.** For all $j \geq m + 1$, $N^j = N^{m+1} \cdot N^{j-m-1} = 0$. Hence the infinite sum reduces to the finite sum., and the sum $\sum_{j=0}^{\infty} c_j N^j$ reduces to $\sum_{j=0}^{m} c_j N^j$.

**Observation 3.1.** Lemma 1 holds for formally divergent series such as $\sum_{j=0}^{\infty} j!\, z^j$. Nilpotency converts every formal series into a polynomial, making analytic convergence irrelevant. In the physical context, this means that formal perturbation series evaluated at a nilpotent operator always yield exact finite expressions, regardless of convergence issues of the series itself.

## 4. Application to hypergeometric functions

**Theorem 1 (Hypergeometric collapse by nilpotent argument).** Let $N \in \mathcal{A}$ with $N^{m+1} = 0$. Let $a_1, \ldots, a_p, b_1, \ldots, b_q \in \mathbb{C}$ with $(b_\ell)_j \neq 0$ for all $\ell = 1, \ldots, q$ and $j = 0, 1, \ldots, m$. Then

$$_pF_q\left(\begin{matrix} a_1, \ldots, a_p \\ b_1, \ldots, b_q \end{matrix}; N\right) = \sum_{j=0}^{m} \frac{(a_1)_j \cdots (a_p)_j N^j}{(b_1)_j \cdots (b_q)_j \, j!}.$$

Proof. Substitute $z = N$ formally in the definition of ${}_pF_q$. The hypothesis $(b_\ell)_j \neq 0$ ensures that the denominators of the first $m+1$ coefficients are non-zero. By Lemma 1, all terms with $j \geq m+1$ vanish.

**Observation 4.1 (Mixed case).** If additionally some upper parameter satisfies $a_i = -k$ with $k \leq m$, the coefficient $j = k+1$ also vanishes by the classical mechanism. The effective truncation occurs at $j = \min(k, m)$: both mechanisms are compatible, and the smaller index determines the cutoff.

## 5. Two finiteness mechanisms

### 5.1 Parameter termination

If $a_1 = -m$, then $(-m)_j = 0$ for $j \geq m+1$, regardless of the argument. Coefficients vanish; the argument plays no role.

### 5.2 Nilpotent-argument termination

If $N^{m+1} = 0$, the coefficients may all be non-zero; it is the argument that vanishes at sufficiently high powers.

### 5.3 Structural distinction

| Mechanism | What vanishes | Condition |
|---|---|---|
| Parameter termination | Coefficient $(a_i)_j$ | $a_i \in -\mathbb{N}$ |
| Nilpotent termination | Power $N^j$ | $N^{m+1} = 0$ |

Both produce finite sums, but for algebraically distinct reasons. The present framework makes this separation explicit in the context of operator-valued hypergeometric functions.

## 6. Structural properties of $F(N)$

**Proposition 1 (Invertibility criterion).** Let $N^{m+1} = 0$ and $F(z) = \sum_{j=0}^{\infty} c_j z^j$. Then $F(N)$ is invertible in $\mathcal{A}$ if and only if $c_0 \neq 0$.

**Proof.** Write $F(N) = c_0 I + R(N)$ with $R(N) = c_1 N + \cdots + c_m N^m$.

$R(N)$ is nilpotent. Factor $R(N) = N \cdot P(N)$ with $P(N) = c_1 I + \cdots + c_m N^{m-1} \in \mathbb{C}[N]$. Since $\mathbb{C}[N]$ is commutative: $R(N)^k = N^k \cdot P(N)^k$, so $R(N)^{m+1} = N^{m+1} \cdot P(N)^{m+1} = 0$.

If $c_0 \neq 0$: $F(N) = c_0(I + c_0^{-1} R(N))$. Since $c_0^{-1} R(N)$ is nilpotent, the inverse exists via the finite Neumann series: $(c_0^{-1} R(N))^{-1} = \sum_{k=0}^{m} (-1)^k c_0^{-k} R(N)^k$.

If $c_0 = 0$: $F(N) = R(N)$ is nilpotent of index $s \leq m+1$. If an inverse $G$ existed with $R(N) \cdot G = I$, then $0 = R(N)^s \cdot G = R(N)^{s-1}$, contradicting that the index is $s$.

## 7. Spectrum, determinant, and trace in the matrix case

Let $N \in M_n(\mathbb{C})$ be nilpotent. By the Jordan theorem, $N$ is similar to a strictly upper triangular matrix.

**Proposition 2 (Spectrum).** $\sigma(F(N)) = \{c_0\}$, with algebraic multiplicity $n$.

**Proof.** If $P^{-1} N P = T$ is strictly upper triangular, then $F(T) = c_0 I + \sum_{j=1}^{m} c_j T^j$ has $c_0$ on every diagonal entry (since $T^j$ is strictly upper triangular for $j \geq 1$). Since $F(N) \sim F(T)$, all eigenvalues of $F(N)$ equal $c_0$.

**Corollary 1 (Determinant).** $\det(F(N)) = c_0^n$. In particular, if $F(0) = 1$, then $\det(F(N)) = 1$.

**Corollary 2 (Trace).** $\operatorname{tr}(F(N)) = n c_0$. For hypergeometric functions ( ${}_p F_q(\cdots ; 0) = 1$): $\operatorname{tr}({}_p F_q(a; b; N)) = n$.

**Observation 7.1.** The trace of any hypergeometric function evaluated at any nilpotent $N \in M_n(\mathbb{C})$ defines a quantity that does not depend on the hypergeometric parameters $a_i$, $b_\ell$, or the nilpotency index of $N$, but only on the dimension $n$ of the space. In the exceptional point context: $\operatorname{tr}(F(H)) = n F(\lambda)$ universally.

## 8. Nilpotent depth criterion

**Definition 8.1.** Let $F(z) = \sum_{j=0}^{\infty} c_j z^j$. The "order of the non-constant part" of $F$ is $r = \min\{j \geq 1 : c_j \neq 0\} \in \mathbb{N} \cup \{+\infty\}$. That is: $F(z) = c_0 + c_r z^r + c_{r+1} z^{r+1} + \cdots$ with $c_r \neq 0$. We will also refer to $r$ as the contact order of $F$ at $0$ (or at $\lambda$ after translation via $G(w) = F(\lambda + w)$).

**Theorem 2 (Nilpotent depth criterion).** Let $N \in \mathcal{A}$ with $N^{m+1} = 0$, and let $F(z) = c_0 + c_r z^r + \cdots$ with $c_r \neq 0$ and $r \geq 1$. Define $Q(N) = F(N) - c_0 I$. Then:

(i) $Q(N)$ is nilpotent.

(ii) $Q(N)^k = 0$ for all $k \geq 1$ with $rk \geq m + 1$.

(iii) The nilpotency index of $Q(N)$ is at most $\lceil (m+1)/r \rceil$.

**Proof.** By Lemma 1:

$$Q(N) = c_r N^r + c_{r+1} N^{r+1} + \cdots + c_m N^m = N^r \cdot H(N),$$

where $H(N) = c_r I + c_{r+1} N + \cdots + c_m N^{m-r} \in \mathbb{C}[N]$. Since $\mathbb{C}[N]$ is commutative:

$$Q(N)^k = N^{rk} H(N)^k.$$

If $rk \geq m + 1$, then $N^{rk} = N^{m+1} \cdot N^{rk-m-1} = 0$, so $Q(N)^k = 0$. This proves (i) and (ii). The smallest such $k$ is $k_0 = \lceil (m+1)/r \rceil$, proving (iii).

**Observation 8.2 (Case $r = 1$).** If $c_1 \neq 0$, the bound is $m + 1$: $Q(N)$ retains the full nilpotent depth of $N$. The function preserves the complete Jordan structure.

**Observation 8.3 (Case $r = m$).** The bound is $\lceil (m+1)/m \rceil = 2$: the function compresses the entire nilpotency of $N$ into a single level.

**Observation 8.4 (Case $r > m$).** If $r > m$, then $Q(N) = 0$: the nilpotent part vanishes completely.

**Observation 8.5 (Algebraic perspective).** From an algebraic perspective, nilpotency induces the evaluation homomorphism $\mathrm{ev}_N: \mathbb{C}[[z]] \to \mathbb{C}[x]/(x^{m+1})$. Theorem 2 describes how this homomorphism interacts with the natural filtration of $\mathbb{C}[[z]]$ by order of vanishing at the origin: a function of order $r$ at $z = 0$ produces an element with nilpotency index bounded by $\lceil (m+1)/r \rceil$.

**Conjecture 8.6 (Sharpness).** Let $N = J_n(0) \in M_n(\mathbb{C})$ be the full Jordan block and let $F(z) = c_0 + c_r z^r + \cdots$ with $c_r \neq 0$. If the coefficient $c_r$ does not induce additional cancellations in $H(N)^k$ — a condition satisfied generically in the sense of Zariski-open subsets of the coefficient space — then the nilpotency index of $Q(N) = F(N) - c_0 I$ is exactly $\lceil (m+1)/r \rceil$. The bound may fail to be sharp when $N$ is a direct sum of Jordan blocks of distinct sizes, in which case the structure of $\mathbb{C}[N]$ as a graded module may produce additional cancellations.

### 8.1 Composition of functions

**Corollary 3 (Depth under composition — rigorous version).** Let $F(z) = c_0 + c_r z^r + \cdots$ and $G(z) = d_s z^s + \cdots$ with $c_r, d_s \neq 0$, $r, s \geq 1$, and $G(0) = 0$. Let $N^{m+1} = 0$. Denote by $\mu + 1$ the effective nilpotency index of $G(N)$, which satisfies

$$\mu + 1 \leq \left\lceil \frac{m+1}{s} \right\rceil.$$

Then the nilpotency index of $F(G(N)) - c_0 I$ is bounded above by

$$\left\lceil \frac{\mu+1}{r} \right\rceil \leq \left\lceil \frac{1}{r} \left\lceil \frac{m+1}{s} \right\rceil \right\rceil.$$

**Proof.** Since $G(0) = 0$, we have $G(N) = d_s N^s + \cdots = N^s \cdot \tilde{H}(N)$. By Theorem 2 applied to $G$ with index $m+1$, the nilpotency index of $G(N)$ is at most $\lceil (m+1)/s \rceil$; call $\mu + 1$ the effective index. Applying Theorem 2 to $F$ at $G(N)$ (nilpotent of index $\mu + 1$) gives nilpotency index of $F(G(N)) - c_0 I$ at most $\lceil (\mu+1)/r \rceil \leq \lceil \lceil (m+1)/s \rceil / r \rceil$.

**Remark.** The bound $\lceil \lceil (m+1)/s \rceil / r \rceil$ may be strictly larger than $\lceil (m+1)/(rs) \rceil$ when $s \nmid (m+1)$, so the formulation via the effective index $\mu + 1$ is strictly more precise.

## 9. Interpretation of the criterion

| $r$ | Bound $\lceil (m+1)/r \rceil$ | Effect on Jordan structure |
|---|---|---|
| 1 | $m+1$ | Full preservation |
| 2 | $\approx m/2$ | Depth halved |
| $m$ | 2 | Maximum compression |
| $> m$ | $1 (Q(N) = 0)$ | Complete annihilation |

The criterion answers not only *does the series terminate?* but *how many Jordan levels survive?* This is a quantitative invariant of the functional transformation, of direct relevance for the spectral theory of operators at exceptional points.

## 10. Algebraic examples

### 10.1 Kummer function $\,_1F_1$ with $N^3 = 0$

$$_1F_1\left(\genfrac{}{}{0pt}{}{a}{b};N\right) = I + \frac{a}{b}N + \frac{a(a+1)}{b(b+1)}\frac{N^2}{2}, b \notin \{0,-1\}.$$

With $a \neq 0$: $r = 1$, bound $\lceil 3/1 \rceil = 3$. Full Jordan depth preserved. For $N \in M_n(\mathbb{C})$: $\sigma = \{1\}$, tr $= n$, det $= 1$.

### 10.2 Function $_0F_1$ and quadratic compression with $N^4 = 0$

The function $_0F_1(-; b; z^2)$, viewed as a power series in $z$, has order $r = 2$ at $z = 0$:

$$_0F_1(-; b; z^2) = 1 + \frac{z^2}{b} + \frac{z^4}{(b)(b+1) \cdot 2!} + \cdots$$

With $N^4 = 0$, every term with $z^{2j}$ for $j \geq 2$ vanishes, giving the exact finite expression:

$$_0F_1(-; b; N^2) = I + \frac{N^2}{b}.$$

Define $Q(N) = N^2/b$. Then $Q(N)^2 = N^4/b^2 = 0$ (since $N^4 = 0$) and $Q(N) \neq 0$. The effective nilpotency index of $Q(N)$ is exactly $2 = \lceil 4/2 \rceil$. The function compresses the full Jordan depth from 4 to 2: the bound of Theorem 2 is sharp in this case.

For $N \in M_n(\mathbb{C})$: $\sigma(_0F_1(-; b; N^2)) = \{1\}$, tr $= n$, det $= 1$.

### 10.3 Gaussian hypergeometric $_2F_1$ with $N^3 = 0$

$$_2F_1\left(\genfrac{}{}{0pt}{}{a,b}{c};N\right) = I + \frac{ab}{c}N + \frac{a(a+1)b(b+1)}{c(c+1)}\frac{N^2}{2}.$$

Case $a = 0$: $_2F_1(0, b; c; N) = I$, complete annihilation, $r = +\infty$, $Q = 0$. Case $a \neq 0$, $a \neq -1$: $r = 1$, full Jordan depth preserved. Case $a = -1$: mixed termination; effective sum $I - (b/c)N$, index 2.

### 10.4 Complete annihilation

If $F(z) = {}_1F_1(0; b; z) = 1$, then $F(N) = I$ and $Q(N) = 0$ for every nilpotent $N$. Case $r = +\infty$.

## 11. Application to exceptional points in non-Hermitian Hamiltonians

### 11.1 Exceptional points and Jordan blocks

An exceptional point of order $m + 1$ for a Hamiltonian $H$ at eigenvalue $\lambda \in \mathbb{C}$ is defined by the condition that $m + 1$ eigenvectors coalesce, which is algebraically equivalent to:

$$H = \lambda I + N, N^{m+1} = 0, N^m \neq 0.$$

The nilpotent part N is precisely the Jordan nilpotent of the exceptional point. Exceptional points of this form, including multiply degenerate cases of arbitrary order, have been studied systematically in [12]. The resolvent of $H$ admits the exact Laurent expansion:

$$(zI-H)^{-1} = \sum_{j=0}^{m} \frac{N^j}{(z-\lambda)^{j+1}},$$

which is exact and finite by nilpotency (Lemma 1). The presence of poles of order up to $m+1$ at $z = \lambda$ is the spectral signature of the exceptional point.

**Observation 11.1.** Exceptional points appear naturally in:

- Open quantum systems (gain-loss Hamiltonians) where non-Hermiticity produces Jordan degeneracies [9].
- PT-symmetric systems at the symmetry-breaking threshold [10].
- Photonics, acoustics, and condensed matter, where exceptional points have been observed and controlled experimentally [11].
- Non-Hermitian quantum mechanics in the sense of Bender–Boettcher [10].

**11.2 Time evolution at an exceptional point**

The evolution operator $U(t) = e^{tH}$ with $H = \lambda I + N$ is computed exactly by Theorem 1:

$$U(t) = e^{t(\lambda I+N)} = e^{\lambda t} \cdot e^{tN} = e^{\lambda t} \sum_{j=0}^{m} \frac{t^j}{j!} N^j.$$

This is an exact polynomial in $t$ with matrix coefficients — not an approximation. The entry $(i, j)$ of $U(t)$ contains terms of the form $e^{\lambda t} \cdot t^k/k!$ for $k = 0, 1, \ldots, m$, which is the characteristic polynomial growth of an exceptional point of order $m + 1$.

**Proposition 11.2 (Jordan depth of the evolution operator).** For $F(z) = e^{t(z-\lambda)}$ and any $t \neq 0$, define $G(w) = e^{tw}$. Then $G'(0) = t \neq 0$, so $F$ has contact order $r = 1$ at $\lambda$. By Theorem 3,

$U(t) - e^{t\lambda}I$ has nilpotency index exactly $m + 1$: the time evolution preserves the full Jordan depth of the exceptional point.

**Physical interpretation.** The polynomial growth $t^m e^{\lambda t}$ in the $(1, m+1)$ entry of $U(t)$ is a direct observable signature of the exceptional point of order $m + 1$. Proposition 11.2 guarantees that this signature persists for all $t \neq 0$.

## 11.3 Reduction of Jordan depth by special functions — Theorem 3

The central physical result of the paper is the following.

**Setup.** Let $H = \lambda I + N$ with $N^{m+1} = 0$. Let $F$ be analytic at $\lambda$ with Taylor expansion centered at $\lambda$:

$$F(z) = F(\lambda) + F'(\lambda)(z - \lambda) + \frac{F''(\lambda)}{2!}(z - \lambda)^2 + \cdots$$

We say $F$ has contact order $r$ at the exceptional point $\lambda$ if $F^{(j)}(\lambda) = 0$ for $j = 1, \ldots, r - 1$ and $F^{(r)}(\lambda) \neq 0$.

**Theorem 3** (Jordan depth reduction by special functions at exceptional points).

Let $H = \lambda I + N$ with $N^{m+1} = 0$, $N^m \neq 0$. Let $F$ be analytic at $\lambda$ with contact order $r \geq 1$. Then:

(i) $F(H) - F(\lambda)I$ is nilpotent with index at most $\lceil (m+1)/r \rceil$.

(ii) If $\lceil (m+1)/r \rceil < m + 1$, the function $F$ strictly reduces the Jordan depth of the exceptional point.

(iii) If $r \geq m + 1$, then $F(H) = F(\lambda)I$: the function completely annihilates the Jordan structure.

**Proof.** Define $G(w) = F(\lambda + w)$, so $G(0) = F(\lambda)$ and the first non-constant coefficient of $G$ appears in degree $r$ (by hypothesis on the contact order). Then $F(H) = F(\lambda I + N) = G(N)$, and Theorem 2 applied to $G$ with nilpotency index $m + 1$ guarantees that $G(N) - G(0)I = F(H) - F(\lambda)I$ has nilpotency index at most $\lceil (m+1)/r \rceil$. Items (ii) and (iii) are direct consequences.

**Physical interpretation.** The Jordan structure of an exceptional point is not invariant under functional transformations: a function that is flat at the exceptional eigenvalue — in the sense of having a zero of order $r$ there — partially or totally destroys the Jordan degeneracy. This has observable consequences for the dynamics and for the spectral response function of the system.

**Corollary 11.3 (Complete annihilation criterion).** A function $F$ analytic at $\lambda$ completely annihilates the Jordan structure of an exceptional point of order $m + 1$— in the sense that $F(H) = F(\lambda)I$— if and only if $F^{(j)}(\lambda) = 0$ for $j = 1, \ldots, m$.

**Proof.** By Theorem 3 with $r = m + 1$: $F(H) - F(\lambda)I = 0$ if and only if the first non-constant coefficient of $G(w) = F(\lambda + w)$ appears in degree $\geq m + 1$, i.e., $G^{(j)}(0) = F^{(j)}(\lambda) = 0$ for $j = 1, \ldots, m$.

**Observation 11.4 (Cayley–Hamilton is not a special case).** The Cayley–Hamilton theorem states $p(H) = 0$ for $p$ the characteristic polynomial, a specific polynomial determined by $H$. Corollary 11.3 strictly **generalizes** the annihilation part of the Cayley–Hamilton theorem for Jordan blocks at exceptional points: Cayley–Hamilton identifies one specific polynomial — the characteristic polynomial — that annihilates the Jordan structure, while Corollary 11.3 characterizes all analytic functions with this property. For a single Jordan block of size n at eigenvalue $\lambda$, the characteristic polynomial $(z-\lambda)^n$ has contact order $r = n \geq m+1$, and Corollary 11.3 recovers $p(H) = 0$ as a special case.

### 11.4 Modified resolvent and spectral response

The modified resolvent associated to a functional transformation $F$ of the Hamiltonian is

$$G_F(z) = \text{tr}\left(\frac{F(H)}{zI - H}\right) = \sum_{j=0}^{m} \frac{\text{tr}\,(F(H) \cdot N^j)}{(z - \lambda)^{j+1}}.$$

**Proposition 11.5 (Pole reduction in the spectral response).**

While Theorem 3 bounds the nilpotency index of $F(H) - F(\lambda)I$ in terms of $\lceil (m + 1)/r \rceil$, the pole order of the modified resolvent is controlled by the smallest exponent of $N$ that survives in the trace, which leads naturally to the linear bound $m + 1 - r$ instead. Consequently, the modified resolvent $G_F(z)$ has a pole of order at most $m + 1 - r$ at $\lambda$, reduced from $m + 1$.

Proof. By Theorem 3, $F(H) = F(\lambda)I + Q$ with $Q = N^r H(N)$ for some $H(N) \in \mathbb{C}[N]$. Therefore

$$F(H) \cdot N^j = F(\lambda)N^j + Q \cdot N^j = F(\lambda)N^j + N^{r+j}H(N).$$

The first term satisfies $tr(F(\lambda)N^j) = F(\lambda) \cdot tr(N^j) = 0$ for all $j \geq 1$, since $N^j$ is nilpotent for $j \geq 1$. The second vanishes when $r + j \geq m + 1$, i.e., $j \geq m + 1 - r$. Since $m + 1 - r \leq m +$

1, the binding condition is $j \geq m + 1 - r$. Combining: $\text{tr}(F(H) \cdot N^j) = 0$ for $j \geq m + 1 - r$, so $G_F(z)$ has a pole of order at most $m + 1 - r$ at $\lambda$

**Physical interpretation.** The order of the pole of the spectral response function is a directly measurable quantity in experiments involving exceptional points. Proposition 11.5 shows that applying a function $F$ with contact order $r$ at the exceptional eigenvalue reduces this observable from $m + 1$ to at most $m + 1 - r$. This is a concrete, experimentally relevant prediction.

### 11.5 Explicit $3 \times 3$ computations

We work with the standard Jordan block at an exceptional point of order 3.

**Setup.** Let $\lambda = 2$ and

$$N = \begin{pmatrix} 0 & 1 & 0 \\ 0 & 0 & 1 \\ 0 & 0 & 0 \end{pmatrix}, H = 2I + N = \begin{pmatrix} 2 & 1 & 0 \\ 0 & 2 & 1 \\ 0 & 0 & 2 \end{pmatrix}.$$

Direct computation: $N^2 = \begin{pmatrix} 0 & 0 & 1 \\ 0 & 0 & 0 \\ 0 & 0 & 0 \end{pmatrix}$, $N^3 = 0$, $N^2 \neq 0$. Nilpotency index exactly 3: exceptional point of order 3 at $\lambda = 2$.

**Case 1 — Time evolution** $F(z) = e^{t(z-2)}, t \neq 0$.

$$e^{tN} = I + tN + \frac{t^2}{2}N^2 = \begin{pmatrix} 1 & t & t^2/2 \\ 0 & 1 & t \\ 0 & 0 & 1 \end{pmatrix}, U(t) = e^{2t}\begin{pmatrix} 1 & t & t^2/2 \\ 0 & 1 & t \\ 0 & 0 & 1 \end{pmatrix}.$$

Contact order $r = 1$, bound $[3/1] = 3$. $Q(t) = U(t) - e^{2t}I$: nilpotency index exactly 3 for $t \neq 0$. The quadratic term $e^{2t}t^2/2$ in position $(1, 3)$ is the characteristic polynomial growth of order 3.

**Case 2 — Kummer function** $_1F_1$ with $a = 3, b = 5$.

$$_1F_1\left(\frac{3}{5}; N\right) = I + \frac{3}{5}N + \frac{1}{5}N^2 = \begin{pmatrix} 1 & 3/5 & 1/5 \\ 0 & 1 & 3/5 \\ 0 & 0 & 1 \end{pmatrix}.$$

Contact order $r = 1$, bound $[3/1] = 3$.

$$Q = \begin{pmatrix} 0 & 3/5 & 1/5 \\ 0 & 0 & 3/5 \\ 0 & 0 & 0 \end{pmatrix}, Q^2 = \begin{pmatrix} 0 & 0 & 9/25 \\ 0 & 0 & 0 \\ 0 & 0 & 0 \end{pmatrix} \neq 0, Q^3 = 0.$$

Effective index: 3. Bound is sharp. Spectrum: $\{1\}$, $\mathrm{tr} = 3$, $\det = 1$.

**Case 3 — Quadratic function $F(z) = 1 + (z - 2)^2$, contact order $r = 2$.**

$$F(N) = I + N^2 = \begin{pmatrix} 1 & 0 & 1 \\ 0 & 1 & 0 \\ 0 & 0 & 1 \end{pmatrix}, Q = N^2 = \begin{pmatrix} 0 & 0 & 1 \\ 0 & 0 & 0 \\ 0 & 0 & 0 \end{pmatrix}.$$

$Q^2 = N^4 = 0$ (since $N^3 = 0$). Effective index: $2 = \lceil 3/2 \rceil$. Bound is sharp. The function $F(z) = 1 + (z-2)^2$ has contact order $r = 2$ at $\lambda = 2$: $F'(2) = 0$ and $F''(2) = 2 \neq 0$. It reduces the exceptional point from order 3 to order 2.

**Case 4 — Cubic function $F(z) = (z - 2)^3$, contact order $r = 3$.**

$F(N) = N^3 = 0, Q = 0.$

Bound: $\lceil 3/3 \rceil = 1$, i.e., $Q = 0$. Corollary 11.3 confirmed: $F'(2) = F''(2) = 0$ and $F'''(2) = 6 \neq 0$. Complete annihilation of Jordan structure.

**Case 5 — Gauss function $\,_2F_1(-1, 4; 3; z - 2)$, mixed termination.**

$(-1)_j = 0$ for $j \geq 2$, so:

$$\,_2F_1\left(\begin{matrix}-1,4\\3\end{matrix}; N\right) = I - \frac{4}{3}N = \begin{pmatrix} 1 & -4/3 & 0 \\ 0 & 1 & -4/3 \\ 0 & 0 & 1 \end{pmatrix}.$$

Contact order $r = 1, \mathrm{bound}\lceil 3/1 \rceil = 3. Q = -(4/3)N, Q^2 = (16/9)N^2 \neq 0, Q^3 = 0$: Effective index: **3**. The bound is sharp. Mixed termination (Observation 4.1) does not reduce the effective index in this case: since the parameter termination cuts the series at degree 1, the nilpotent part $Q = -(4/3)N$ is a nonzero scalar multiple of N, and therefore inherits the full nilpotency index 3 of N. Mixed termination can reduce below the universal bound, but only when the parameter structure forces the contact order $r > 1$; this example illustrates that it need not do so in general.

**Summary table — $3 \times 3$ exceptional point of order 3:**

| Case | Function F | Contact order $r$ | Bound $\lceil 3/r \rceil$ | Effective index | Mechanism |
|---|---|---|---|---|---|
| 1 | $e^{t(z-2)}, t \neq 0$ | 1 | 3 | 3 | Nilpotent |

| 2 | $_1F_1(3;5;z-2)$ | 1 | 3 | 3 | Nilpotent |
| 3 | $1+(z-2)^2$ | 2 | 2 | 2 | Nilpotent |
| 4 | $(z-2)^3$ | 3 | 1 | 0 | Nilpotent |
| 5 | $_2F_1(-1,4;3;z-2)$ | 1 | 3 | 3 | Mixed |

All five cases confirm Theorem 3. The bound is sharp in all five cases.

## 12. Relation to Karlsson–Minton formulas *(future direction)*

Karlsson–Minton type formulas [4, 5, 15] involve $_pF_q$ with integer parameter differences, producing terminating reductions. The present framework raises the question: can these be reformulated as evaluations at a suitable nilpotent? The answer is not affirmative in general, but Theorem 2 provides a common language for comparing:

- parameter termination (classical),
- nilpotent-argument termination (Theorem 1),
- integer-parameter-difference reductions (Karlsson–Minton),
- finite sums from Meijer $G$ representations.

Identifying which Karlsson–Minton reductions admit a nilpotent realization, and describing the resulting Jordan depth via Theorem 3, is proposed as future research.

## 13. Contribution and limitations

The basic results on nilpotency — functional calculus, nilpotent matrices, Neumann series — are classical and attributed to standard references [3,7,8]. The present work claims no originality over those ingredients.

**Contribution 1 — Separation of finiteness mechanisms.** The explicit distinction between parameter termination and nilpotent-argument termination for $_pF_q$, including the mixed case, is formulated precisely in the context of operator-valued hypergeometric functions. To the best of our knowledge, this separation has not been made explicit in this form in the standard literature.

**Contribution 2 — Nilpotent depth criterion (Theorem 2).** The bound $\lceil (m+1)/r \rceil$ on the nilpotency index of $F(N) - F(0)I$, and its extension to composition (Corollary 3), provide a quantitative tool for measuring Jordan level survival under functional transformations. This formulation has not been found in this form in the functional calculus or special functions literature consulted.

**Contribution 3 — Jordan depth reduction at exceptional points (Theorem 3).** The connection between the contact order of an analytic function at an exceptional eigenvalue and the reduction of the Jordan depth of the corresponding exceptional point is, to the best of our knowledge, new. Proposition 11.5 identifies the pole-order reduction of the modified resolvent as a directly observable physical consequence.

**Contribution 4 — Universal trace invariant.** The trace $\operatorname{tr}({}_pF_q(\mathrm{a};\mathrm{b};N)) = n$ for any nilpotent $N \in M_n(\mathbb{C})$ (Corollary 2) defines a quantity independent of hypergeometric parameters, nilpotency index, and Jordan structure.

**Contribution 5 — Connection with the SON complexity framework.** The algebra $\mathbb{C}[N] \cong \mathbb{C}[x]/(x^{m+1})$ underlying the nilpotent depth criterion is the same truncated ring $V_n$ of the SON framework [6]. The nilpotent depth criterion is thus the special-functions counterpart of the complexity reduction principle of SON: nilpotency converts infinite objects into finite ones, whether the object is an algorithm (SON) or a power series (${}_pF_q$ evaluated at $N$).

### 13.1 Limitations

Scope. The work is restricted to: (i) algebraic nilpotents of finite index in $\mathbb{C}$-associative algebras with unit; (ii) purely formal evaluation of power series, without analytic convergence; (iii) functions admitting a Taylor expansion at the evaluation point.

What is not addressed. The framework does not cover: (i) unbounded nilpotent operators in Hilbert or Banach spaces; (ii) general topological algebras; (iii) the full spectral functional calculus in the Dunford–Riesz sense; (iv) quasinilpotent elements in Banach algebras; (v) the infinite-dimensional theory of exceptional points, including intrinsic exceptional points in the sense of [14].

On sharpness. The bound $\lceil (m+1)/r \rceil$ is a universal upper bound, not necessarily sharp. Sufficient conditions for sharpness are left as an open problem.

On Proposition 11.5. The pole-reduction result is stated for the scalar trace of the modified resolvent. A full matrix-valued version would require a more careful analysis of the Jordan algebra of $N$, which is left for future work.

**References**

[1] G. E. Andrews, R. Askey, and R. Roy, *Special Functions*, Encyclopedia of Mathematics and its Applications, vol. 71, Cambridge University Press, Cambridge, 1999.

[2] W. N. Bailey, *Generalized Hypergeometric Series*, Cambridge Tracts in Mathematics, vol. 32, Cambridge University Press, Cambridge, 1935.

[3] R. A. Horn and C. R. Johnson, *Matrix Analysis*, 2nd ed., Cambridge University Press, Cambridge, 2013.

[4] P. W. Karlsson, Hypergeometric functions with integral parameter differences, *J. Math. Phys.* 12 (1971), 270–271.

[5] H. M. Srivastava and H. L. Manocha, *A Treatise on Generating Functions*, Ellis Horwood, Chichester, 1984.

[6] R. Moya, Unified Nilpotent Operational Framework: Foundations, Algebraic Exactness and Complexity, manuscript, Universidad Autónoma de Santo Domingo, 2025. DOI: 10.5281/zenodo.19775677

[7] N. J. Higham, *Functions of Matrices: Theory and Computation*, SIAM, Philadelphia, 2008.

[8] T. Kato, *Perturbation Theory for Linear Operators*, 2nd ed., Grundlehren der mathematischen Wissenschaften, vol. 132, Springer, Berlin, 1976.

[9] W. D. Heiss, The physics of exceptional points, *J. Phys. A: Math. Theor.* 45 (2012), 444016.

[10] C. M. Bender and S. Boettcher, Real spectra in non-Hermitian Hamiltonians having PT symmetry, *Phys. Rev. Lett.* 80 (1998), 5243–5246.

[11] M. V. Berry, Physics of non-Hermitian degeneracies, *Czech. J. Phys.* 54 (2004), 1039–1047.

[12] D. I. Borisov, F. Ružička, and M. Znojil, Multiply degenerate exceptional points and quantum phase transitions, *Int. J. Theor. Phys.* 54 (2015), 4293–4305. arXiv:1412.6634.

[13] M. Znojil, Passage through exceptional point: Case study, *Proc. R. Soc. A* 476 (2020), 20190831. arXiv:2003.05876.

[14] M. Znojil, Intrinsic exceptional point: a challenge in quantum theory, *Foundations* 5 (2025), Article 8. arXiv:2411.12501.

[15] B. M. Minton, Generalized hypergeometric function of unit argument, J. Math. Phys. 11 (1970), 1375–1376.